\documentclass[aps, prd, reprint, twocolumn, tightenlines, superscriptaddress, amsmath, amssymb, nofootinbib]{revtex4-1}
\usepackage{color}
\usepackage{amsmath}
\usepackage{graphicx}
\DeclareGraphicsExtensions{.pdf,.png,.jpg}

\begin{document}

%%%%%%%%%%%%%%%%%%%%%%%%%%%%%%%%%%%%%%%%%%%%%%%%%%%%%%%%%%
\title{Dark Energy Induced Anisotropy in Cosmic Expansion}
%%%%%%%%%%%%%%%%%%%%%%%%%%%%%%%%%%%%%%%%%%%%%%%%%%%%%%%%%%

\author{Chien-Ting Chen}
\email{f98244003@ntu.edu.tw}             

\affiliation{Department of Physics, National Taiwan University, Taipei, Taiwan 10617}
\affiliation{Leung Center for Cosmology and Particle Astrophysics, National Taiwan University, Taipei,
             Taiwan 10617}
\author{Pisin Chen}
\email{pisinchen@phys.ntu.edu.tw}
\affiliation{Department of Physics, National Taiwan University, Taipei, Taiwan 10617}
\affiliation{Leung Center for Cosmology and Particle Astrophysics, National Taiwan University, Taipei,
             Taiwan 10617}
             \affiliation{Graduate Institute of Astrophysics, National Taiwan University, Taipei, Taiwan 10617}
\affiliation{Kavli Institute for Particle Astrophysics and Cosmology, SLAC National Accelerator Laboratory,
             Stanford University, Stanford, CA 94305, U.S.A.}

\begin{abstract}
In order to understand the nature of the accelerating expansion of the late-time universe, it is important to experimentally determine whether dark energy is a cosmological constant or dynamical in nature. If dark energy already exists prior to inflation, which is a reasonable assumption, then one expects that a dynamical dark energy would leave some footprint in the anisotropy of the late-time accelerated expansion.
To demonstrate the viability of this notion, we invoke the quintessence field with the exponential potential as one of the simplest dynamical dark energy models allowed by observations. We investigate the effects of its quantum fluctuations (the physical origin of the perturbation being isocurvature) generated during inflation and having fully positive correlation with the primordial curvature perturbations, and estimate the anisotropy of the cosmic expansion so induced. We show that the primordial amplitude of quantum fluctuations of quintessence field $ \delta \phi_{\textrm{P}} $ can be related to the tensor-to-scalar ratio $ r　$, and we calculate the perturbed luminosity distance to first order and the associated luminosity distance power spectrum, which is an estimator of anisotropicity of late-time accelerated expansion. 
We find that the gravitational potential at large scales and late times is less decayed in QCDM compared to that in $ \Lambda $CDM so that the smaller the redshift and multipole, the more relative deficit of power in QCDM. 
Our results of luminosity distance power spectrum also show the similar conclusions of suppression as that of the previous investigation regarding the effect of quantum fluctuations of quintessence field on the CMB temperature anisotropies.
\end{abstract}

\maketitle

%%%%%%%%%%%%%%%%%%%%%%%%%%%%%%%%%%
\section{Introduction}\label{sec1}
%%%%%%%%%%%%%%%%%%%%%%%%%%%%%%%%%%

We now believe that our universe is currently undergoing a phase of late-time accelerated expansion through the observations of supernovae \cite{Perlmutter:1996ds, Perlmutter:1997zf, Perlmutter:1998np, Riess:1998cb, Riess:1998dv, Riess:2001gk, Riess:2004nr, Riess:2006fw, Kowalski:2008ez, Hicken:2009dk, Lampeitl:2009jq, Amanullah:2010vv, Rest:2013mwz, Betoule:2014frx}. To explain this phenomenon, one generally assumes the existence of a substance called dark energy that constitutes roughly 70$ \% $ of the total energy density of the present day universe. 
The simplest candidate for dark energy is the cosmological constant $ \Lambda$, based on which the $\Lambda$CDM model has been proven consistent with various observations such as baryon acoustic oscillations (BAO) \cite{Eisenstein:2005su, Benitez:2008fs} and cosmic microwave background (CMB) \cite{Hinshaw:2012aka, Ade:2015xua}. 
However, the $ \Lambda$CDM model suffers from two unexplainable conceptual problems, namely the ``fine tuning problem", which concerns why the energy density of $\Lambda$ is so tiny compared with the vacuum energy density expected from quantum field theory, and the ``coincidence problem", which wonders why the energy density of $\Lambda$ is comparable to the matter density of today \cite{Zlatev:1998tr}. 
Alternative candidates for dark energy, which may or may not be motivated by solving either of these two problems, typically invoke a dynamical scalar field, among them one simple construction is the quintessence field $ \phi $ \cite{Caldwell:1997ii, Huey:1998se, Steinhardt:1999nw, Brax:1999yv}.

Thus far there is no obvious logical connection between the emergence of dark energy and any particular era in the cosmic evolution. It is therefore customary to assume that the dark energy already exists at the beginning of time. If so, then since the quintessence is a dynamical scalar field, it must induce quantum fluctuations during inflation and may therefore leave imprints on cosmological observables such as the CMB, the late-time accelerating expansion, etc. In contrast, the cosmological constant is not expected to induce such quantum fluctuations.
In \cite{Kawasaki:2001bq, Kawasaki:2001nx, Moroi:2003pq}, the issue about the effect of quantum fluctuations of quintessence field on the CMB temperature anisotropies has been investigated, and it has been shown that if the amplitude of primordial quantum fluctuations of the quintessence field $ \delta \phi_{\textrm{P}} $ has fully positive correlation with the primordial amplitude of the curvature perturbation $ \Phi_{\textrm{P}} \approx 10^{-5} $, then there should be a sizable suppression in the low multipoles of CMB temperature anisotropies relative to that predicted by the $ \Lambda $CDM model, while in the uncorrelated case the outcome is opposite.
In this work, we consider quantum fluctuations of the quintessence field during inflation as the physical origin of a possible anisotropy of the late time accelerating expansion. Unlike most previous investigations, where the seeds of such late time expansion anisotropy were put in by hand in an ad hoc way, here we provide a self-consistent ab initio mechanism to anchor the issue on a firmer basis.

There also exists tentative evidence of a preferred axis in the supernovae distribution, which suggests that the late-time accelerating expansion may be anisotropic \cite{Kolatt:2000yg, Schwarz:2007wf, Kalus:2012zu, Campanelli:2010zx, Mariano:2012wx, Cai:2013lja, Cooke:2009ws, Antoniou:2010gw, Heneka:2013hka, Zhao:2013yaa, Yang:2013gea, Javanmardi:2015sfa, Lin:2015rza}, though not yet statistically significant. In addition, there are hints on preferred axes in the observed bulk velocity flows, the alignment of CMB low multipoles (including dipole, quadrupole and octupole), and that of the quasi-stellar object (QSO) optical polarization \cite{Antoniou:2010gw, Heneka:2013hka, Zhao:2013yaa, Salehi:2016sta, Zhao:2016fas}. These studies show that the coexistence of these independent cosmological axes in a relatively small angular region is statistically unlikely, but more probably induced by either an undiscovered physical effect or a systematic error that has so far escaped attention.
Dark energy induced spatial inhomogeneities has been investigated phenomenologically \cite{Cooray:2008qn, Blomqvist:2010ky}. Here we point out that such anisotropy is a necessary consequence of dynamical dark energy if it already exists prior to the inflation era.

For the purpose of illustrating our point, we invoke the quintessence field $ \phi $ with the exponential potential, which is one of the simplest dynamical dark energy models allowed by observations, and consider the effects of ``quantum fluctuations of quintessence'' on the late time cosmic expansion. The evolution of the universe led by this field will be described in Sec. \ref{sec2}. The gravitational potential originated from the scalar perturbations will affect the trajectory of light rays. In Sec. \ref{sec3}, we calculate the perturbed luminosity distance resulting from the evolution of the perturbed gravitational potential \cite{Sasaki:1987ad, Pyne:1993np, Pyne:1995ng, Pyne:2003bn, Bonvin:2005ps, Hui:2005nm, Flanagan:2008kz}. We then calculate the corresponding power spectrum to estimate the anisotropy of the late time accelerating expansion \cite{Bonvin:2005ps}. The numerical results for the evolution of the universe as well as the luminosity distance power spectrum will be shown in Sec. \ref{sec4}. Finally, the conclusion will be given at the end of this paper.

%%%%%%%%%%%%%%%%%%%%%%%%%%%%%%%%%%%%%%%%%%%%%%%%%
\section{The Quintessential Universe}\label{sec2}
%%%%%%%%%%%%%%%%%%%%%%%%%%%%%%%%%%%%%%%%%%%%%%%%%

Let us describe the universe via the QCDM model, composed of the quintessence field (Q) and the pressureless matter $m$ including the cold dark matter (CDM) and the baryonic matter $b$, while other species are neglected. Throughout this paper, we adopt the exponential potential for the quintessence field $ \phi $ of the form:
\begin{equation}\label{frw}
   V(\bar{\phi}) = V_0 \exp \left( - \lambda \frac{\bar{\phi}(\eta)}{M_\text{P}} \right), 
\end{equation}
where $ \bar{\phi} $ is the background value of the quintessence field, $ \lambda$ and $ V_0 $ are the two parameters, $ M_\text{P} = 1 / \sqrt{8 \pi G} $ is the reduced Planck mass, and we focus on the evolution of the universe from the matter-dominant era, well after the matter-radiation equality, to the present epoch.

%%%%%%%%%%%%%%%%%%%%%%%%%%%%%%%%%%%%%%%%%%%%%%%
\subsection{Background Evolution}\label{sec2.1}
%%%%%%%%%%%%%%%%%%%%%%%%%%%%%%%%%%%%%%%%%%%%%%%

The Friedman-Robertson-Walker (FRW) metric for the spatially flat homogeneous and isotropic universe is given by
\begin{equation}\label{frw}
  ds^2 = a^2(\eta) \left( - d\eta^2 + \delta_{ij} dx^i dx^j \right), 
\end{equation}
where $ \eta $ is the conformal time and $ a(\eta) $ is the scale factor normalized to $ 1 $ at the present time $ \eta_0 $. 

Since the quintessence is a quantum scalar field, its background energy density and pressure are
\begin{subequations}
\begin{align}
  \displaystyle
  \rho_{\phi}(\eta) = & \frac{1}{2 a^2(\eta)} \bar{\phi}'^2 + V(\bar{\phi}), \\
  \displaystyle
  p_{\phi}(\eta) = & \frac{1}{2 a^2(\eta)} \bar{\phi}'^2 - V(\bar{\phi}),
\end{align}
\end{subequations}
where the prime denotes the derivative with respect to the conformal time $ \eta $.
The background evolution of the universe, i.e., the scale factor $ a(\eta) $, is governed by the background Friedmann equation 
\begin{align}\label{bgf}
  \displaystyle 
  \mathcal{H}^2 = & \frac{8 \pi G a^2}{3} 
                    \left( \rho_m + \rho_{\phi} \right),
\end{align}
where $ \mathcal{H} \equiv a' / a $ is the Hubble parameter expressed in the conformal time. The background equation of motion for the quintessence field $ \phi $ is given by the background Klein-Gordon equation 
\begin{equation}\label{bgk}
  \displaystyle 
  \bar{\phi}'' + 2 \mathcal{H} \bar{\phi}' + a^2 V,_{\bar{\phi}}                                                                = 0,
\end{equation}
where $ ,_{\bar{\phi}} $ denotes the derivative with respect to $ \bar{\phi} $.
Solving the coupled equations \eqref{bgf} and \eqref{bgk} together, we obtain the evolution of the scalar field and the quintessence field $ \bar{\phi}(\eta) $.

%%%%%%%%%%%%%%%%%%%%%%%%%%%%%%%%%%%%%%%%%%%%%%%%%%%%%%%%%%%%
\subsection{Evolution of Scalar Perturbations}\label{sec2.2}
%%%%%%%%%%%%%%%%%%%%%%%%%%%%%%%%%%%%%%%%%%%%%%%%%%%%%%%%%%%%

The line element of the perturbed FRW metric in the conformal Newtonian gauge with the case of zero proper anisotropy of the medium is given by \cite{Kodama:1985bj}
\begin{equation}\label{ppfrw}
  ds^2 = a^2(\eta) \left[ ( 1 + 2\Phi ) d\eta^2 - ( 1 - 2\Phi ) \delta_{ij} dx^i dx^j \right],
\end{equation}
where $ \Phi(x, \eta) $ is a gauge-invariant metric perturbation called the Bardeen potential \cite{Bardeen:1980kt}, equivalent to the gravitational potential, and $ a(\eta) $ is the scale factor. 
To first order in perturbations, the quintessence field is decomposed as $ \phi = \bar{\phi} + \delta \phi $, where $ \delta \phi $ is the quantum fluctuating part of $ \phi $. The dynamics of $ \delta \phi $ is governed by the background Klein-Gordon equation \eqref{bgk} and the perturbed Klein-Gordon equation \cite{Riotto:2002yw}
\begin{align}\label{pk}
  \displaystyle
  \delta\phi'' & + 2 \mathcal{H} \delta\phi' + k^2 \delta\phi - 4 \Phi' \bar{\phi}' 
    + a^2 \left[ 2 \Phi V,_{\bar{\phi}} - V,_{\bar{\phi}\bar{\phi}} \delta\phi
          \right] = 0,
\end{align}

The linearized Einstein equations \cite{Riotto:2002yw} with respect to the metric of Eq. \eqref{ppfrw} compose a system involving pressureless matter of density $ \rho_m $ and quintessence field $ \phi $ (with density and pressure perturbations $\delta \rho_{\phi}$ and $\delta p_{\phi}$). This sytem is given, in Fourier space, by
\begin{align}
\label{le00}
  \displaystyle 
  3 \mathcal{H} \Phi' + 3 \mathcal{H}^2 \Phi + k^2 \Phi 
    = - 4 \pi G \left( \delta \rho_m + \delta \rho_{\phi} \right) , \\[0.7em]
\label{le}
  \displaystyle 
  \Phi'' + 3 \mathcal{H} \Phi' + \left( 2 \frac{a''}{a} - \mathcal{H}^2 \right) \Phi 
    = 4 \pi G \, \delta p_{\phi} , \\[0.7em]
  \displaystyle 
  \Phi' +  \mathcal{H} \Phi 
    = - 4 \pi G  \left( \rho_m v_m + \rho_{\phi} v_{\phi} \right) , \\[0.7em]
 \displaystyle 
 \left\lbrace \begin{array}{c}
     \delta \rho_{\phi} \\ 
     \delta p_{\phi}
   \end{array} \right\rbrace 
   \equiv \bar{\phi}' \delta\phi' - \Phi \bar{\phi}'^2 \pm V,_{\bar{\phi}} \delta\phi.
\label{rhop}
\end{align}
Since matter is pressureless, the fluctuations in pressure are only sourced by the quintessence field. Therefore, it is convenient to choose Eq. \eqref{le}, together with Eq. \eqref{pk}, to solve the coupled equations of perturbations and obtain the solutions for the evolution of the gravitational potential $ \Phi_k $ by which the photon trajectories will be affected. 
It is also convenient to normalize $ \Phi_k $ with respect to its initial value $ \Phi_i \equiv \Phi(\eta_i) $, where $ \eta_i $ denotes the initial time of our period of interest, by $ \Phi_{Nk} \equiv \Phi_k / \Phi_i $. For small scales that reentered the horizon before matter-radiation equality, their gravitational potential has already decayed. Therefore, the correct evolution of gravitational potential is given by $ \Phi_{Nk} $ during matter-dominant era weighted by the transfer function $ T(k) $. This function can be approximated by $ T^2(k) \simeq 1 / [1 + \beta (k \eta_{\text{eq}})^4] $, where $ \eta_{\text{eq}} $ denotes the value of conformal time at matter-radiation equality and $ \beta \simeq 3 \times 10^{-4} $ \cite{Bonvin:2005ps}.

%%%%%%%%%%%%%%%%%%%%%%%%%%%%%%%%%%%%%%%%%%%%%%%%%%%%%%%%%%%%%%%%%%%%%
\subsection{Quantum Fluctuations of Quintessence Field}\label{sec2.3}
%%%%%%%%%%%%%%%%%%%%%%%%%%%%%%%%%%%%%%%%%%%%%%%%%%%%%%%%%%%%%%%%%%%%%

To solve the coupled equations, \eqref{pk} and \eqref{le}, it is necessary to find the appropriate initial conditions for $ \delta \phi $. We make two assumptions for the quintessence field. First, the quintessence field already exists during inflation, whose energy density is much smaller than that of the inflaton field, and hence the expansion rate of the universe during inflation, $ H_{\textrm{inf}} $, is governed by inflaton field alone.
Second, after inflation, the quintessence field is still subdominant in radiation- and matter-dominant eras and become dominant only at late times, ushering in the dark energy-dominant era. This is in effect a two-scalar-field inflation scenario with the total energy density being dominated by the inflaton field during inflationary era. Consequently, the primordial quantum fluctuations of the quintessence field considered in this work are isocurvature perturbatons \cite{Kawasaki:2001bq, Kawasaki:2001nx, Moroi:2003pq}.

In this scenario, each mode of quantum fluctuations of quintessence field will be stretched to outside the horizon and become frozen due to the rapid expansion, and therefore turn classical. If the quintessence field has a very light mass compared to the expansion rate during inflation, i.e., $ m_{\phi} \ll H_{\textrm{inf}} $, which is a typical assumption, then the amplitude of its quantum fluctuations is \cite{Kawasaki:2001bq, Riotto:2002yw}
\begin{equation}\label{dphii}
  \delta \phi = \frac{H_{\textrm{inf}}}{2\pi}.
\end{equation}
For simplicity, we also assume that this amplitude \eqref{dphii} is scale-invariant since the only source of scale dependence is the mild variation of $ H_{\textrm{inf}} $ during inflation, which, for our purpose, can be ignored.

After inflation, each fluctuation mode of quintessence field will reenter the horizon and will thaw and evolve, starting from the primordial amplitude
\begin{equation}\label{dphip}
  \delta \phi_\text{p} = \frac{H_{\textrm{inf}}}{2\pi}.
\end{equation}
The primordial amplitude \eqref{dphip} can be parameterized with respect to the amplitude of primordial curvature perturbation $ \Phi_\text{p} $ by\footnote{In fact, there are two types of definition for the parameterization $ \delta y_\text{p} $ as described in \cite{Moroi:2003pq}. If $ \delta \phi_\text{p} $ and $ \Phi_\text{p} $ are fully correlated, then $ \delta y_\text{p} \equiv \delta \phi_\text{p} / M_\text{P} \Phi_\text{p} $. However, if $ \delta \phi_\text{p} $ and $ \Phi_\text{p} $ are uncorrelated, the parameterization has to be as $ \delta y_\text{p} \equiv \sqrt{\langle \delta \phi^2_\text{p} \rangle} / M_\text{P} \sqrt{\langle \Phi^2_\text{P} \rangle} $. We only consider the case of full correlation in this work.}
\begin{equation}
  \delta y_\text{p} \equiv \frac{\delta \phi_\text{p}}{M_\text{P} \Phi_\text{p}},
\end{equation}
advantageously related to the tensor-to-scalar ratio $ r $ which we can show below. Through the analysis of primordial tensor perturbations \cite{Planck:2013jfk, Ade:2015lrj}, the amplitude of tensor power spectrum $ A_\text{T} $ is given by $ A_\text{T} = A_{\Phi} \, r = \Phi^2_\text{p} \, r  \approx 2 V_{\textrm{inf}} / 3 \pi^2 M^4_\text{P} $, where the amplitude of curvature power spectrum $ A_{\Phi} $ is exactly equal to $ \Phi^2_\text{p} $, and  $ V_{\textrm{inf}} $ is the potential of inflaton field. During inflation and under slow-roll approximation, the Hubble parameter $ H_{\textrm{inf}} $ can be written as $ H^2_{\textrm{inf}} \approx V_{\textrm{inf}} / 3 M^2_\text{P} $. After some substitutions, one obtains a relation between the $ \delta \phi_\text{p} $ and $ r $ as 
\begin{equation}
  \delta y_\text{p} \approx \sqrt{\frac{r}{8}},
\end{equation}
which indicates two meanings. First, since $ \delta y_\text{p} $ is positive in our numerical analysis, the primordial amplitude of quantum fluctuations of the quintessence field, $ \delta \phi_\text{p} $, has positive correlation with the primordial curvature perturbation, $ \Phi_\text{p} $. Second, since there is no phase difference between $ \delta \phi_\text{p} $ and $ \Phi_\text{p} $ in our parameterization, $ \delta \phi_\text{p} $ and $ \Phi_\text{p} $ are fully correlated with each other as described in \cite{Moroi:2003pq}.

The initial value $ \delta \phi_i \equiv \delta \phi(\eta_i) $ in Eqs. \eqref{pk} and \eqref{le} can be obtained through $ \delta \phi_\text{p} $ evolving from the end of inflation to the time $ \eta_i $. 
Since most of the modes of interest, concerning the observables to be investigated at $ z \leq 0.5 $ and $ \ell \leq 10 $, cross the horizon long after matter-radiation equality, the gravitational potential for corresponding modes is constant at both radiation- and matter-dominant eras (leading to $ \Phi' = 0 $ at these times in Eq. \eqref{pk}). Along with the fact that $ \phi $ is subdominant and $ \delta \phi $ is solely determined by $ \Phi $, one obtains that $ \delta \phi $, following the same behavior of $ \Phi $, is also constant for both eras and finally we conclude that $ \delta \phi_i = T(k) \delta \phi_{\text{p}} $.

%%%%%%%%%%%%%%%%%%%%%%%%%%%%%%%%%%%%%%%%%%%%%%%%%%%%%%%%%%%%%%%%%%%%%%%%%%%%%%%%%%%%%%%%%%%%%%
\section{The perturbed Luminosity Distance and Luminosity Distance Power Spectrum}\label{sec3}
%%%%%%%%%%%%%%%%%%%%%%%%%%%%%%%%%%%%%%%%%%%%%%%%%%%%%%%%%%%%%%%%%%%%%%%%%%%%%%%%%%%%%%%%%%%%%%

The gravitational potential $ \Phi(k, \eta) $ is constant in time during the matter-dominant era, while it decays during the dark energy dominant era due to the negative pressure of dark energy. This is true no matter whether the dark energy is dynamical or a cosmological constant. Furthermore, if dark energy is indeed the cosmological constant $ \Lambda $, then the amount of decay for $ \Phi(k, \eta) $ will be uniform for all scales because there would be no fluctuations from $ \Lambda $. On the other hand, if dark energy is a quintessence field $ \phi $, then due to its mode-dependent fluctuations $ \delta \phi $, the amount of decay for $ \Phi(k, \eta) $ will be different depending on scales and directions. Hence the propagation of photons varies in different parts of the sky. The luminosity distance $ d_L $ therefore depends not only on the redshift $ z $ but also on the direction $ \mathbf{n} $ in the sky, i.e., $d_L= d_L(z, \mathbf{n})$, discriminating in principle the quintessence field from the cosmological constant.

Consider a standard candle with four-velocity $ u_S $ at a spacetime position $ S $ that emits a total luminosity $ L $,  which is received with an energy flux $F$ by an observer with four-velocity $ u_O $ at a spacetime position $ O $. The luminosity distance between $ S $ and $ O $, $ d_L(S,O) = \sqrt{ L / 4 \pi F}　$, can be obtained by knowing the redshift $ z_S $, the direction $ \mathbf{n} $, and the intrinsic luminosity of the source, and by measuring its flux $ F $.

When there are fluctuations generated by the constituents of the universe, the spacetime geometry is distorted from its unperturbed configuration due to the time evolution and the spatial variation of the gravitational potential $ \Phi(x, \eta) $ so induced, and the path of photons is altered accordingly, differently from the path in the background universe.
To first order in scalar perturbations, we consider the fully perturbed luminosity distance $ d_L(z_S,\mathbf{n}) $ of a source at redshift $ z_S $ and radial direction $ \mathbf{n} $, with the corresponding luminosity distance power spectrum formulated in \cite{Bonvin:2005ps}. An equivalent formula for $ d_L(z_S,\mathbf{n}) $ can be found in \cite{Pyne:2003bn, Hui:2005nm, Flanagan:2008kz}. Formulas that retain second order perturbations are also available in the literature \cite{BenDayan:2012wi, Nugier:2013tca}. The angular power spectrum $ C_{\ell} $ is defined by $ C_{\ell}(z_S, z_{S'}) = \langle a_{\ell m}(z_S) a^*_{\ell m}(z_{S'}) \rangle $, with $ d_L(z_S,\mathbf{n}) = \sum_{\ell m} a_{\ell m}(z_S)Y_{\ell m}(\mathbf{n}) $.
The correlation function of luminosity distance can be obtained by applying the addition theorem for spherical harmonics as 
\begin{align}\label{cfldk}
  \displaystyle
  \frac{\langle d_L(z_S,\mathbf{n}) d_L(z_{S'},\mathbf{n}') \rangle}{\bar{d}_L(z_S) \bar{d}_L(z_{S'})}
  =  \sum_{\ell} \frac{2\ell + 1}{4\pi} P_{\ell}(\mathbf{n}, \mathbf{n}') 
      \sum_{i = 1}^5 C^{(i)}_{\ell},
\end{align}
where $ P_{\ell} $ is the Legendre polynomial of order $ \ell $ and $ C^{(i)}_{\ell} =\int dk k^{i-2}C_{\ell}  $ collects all the contributions of $C_{\ell}$ through the integration \cite{Bonvin:2005ps}. For details of the fully perturbed luminosity distance $ d_L(z_S,\mathbf{n}) $ to first order and the components $ C^{(i)}_{\ell} $ of the corresponding luminosity distance power spectrum formulated in \cite{Bonvin:2005ps} see Appendices A and B.
In this work, we calculate the luminosity distance power spectrum for the QCDM model, accounting for the complexity of the evolution of the gravitational potential described by Eqs. \eqref{pk} and \eqref{le}, and concentrate on the case where $ z_S = z_{S'} $ for simplicity.

%%%%%%%%%%%%%%%%%%%%%%%%%%%%%%%%%%%%%%%
\section{Numerical Results}\label{sec4}
%%%%%%%%%%%%%%%%%%%%%%%%%%%%%%%%%%%%%%%

Throughout the numerical calculations, we adopt the data from Planck mission  \cite{Aghanim:2018eyx, Akrami:2018odb} as the fiducial values to fix $ \lambda = 0.5 $ in the potential $ V(\phi) $ and take the temporal range to be from $ z = 100 $ (i.e. in the matter-dominated era well after the matter-radiation equality) to the present time.

%%%%%%%%%%%%%%%%%%%%%%%%%%%%%%%%%%%%%%%%%%%%%%%
\subsection{Background Evolution}\label{sec4.1}
%%%%%%%%%%%%%%%%%%%%%%%%%%%%%%%%%%%%%%%%%%%%%%%

\begin{figure}[t]
 \centering
 \includegraphics[width=\linewidth]{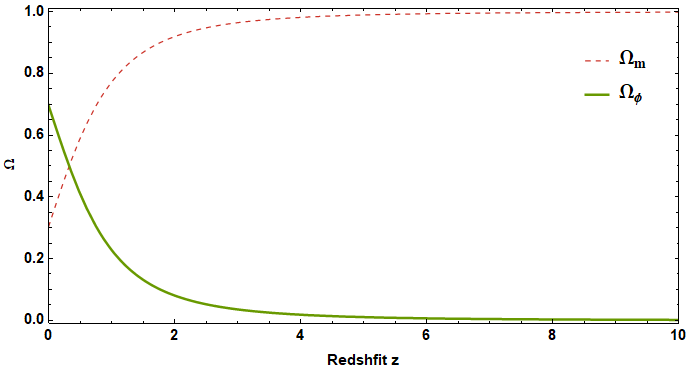}
 \caption{The evolution of the density parameters of matter $ \Omega_m $ (dashed red) and of quintessence field $ \Omega_{\phi} $ (solid green).\label{omega}} 
\end{figure}
\begin{figure}[t]
 \centering
 \includegraphics[width=\linewidth]{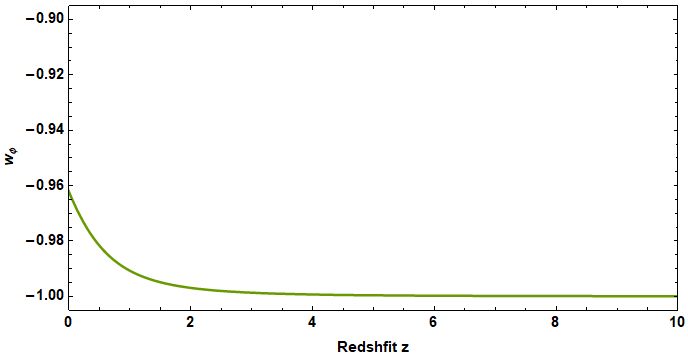}
 \caption{The equation of state of the quintessence field $ w_{\phi} $. When $ z = 0 $, $ w_{\phi} \approx -0.96 $.\label{eos}}
\end{figure}

By solving the coupled equations \eqref{bgf} and \eqref{bgk}, we obtain the background evolution of the universe, which can be demonstrated by the evolution of the density parameters shown in Fig. \ref{omega}. In this figure, the density parameter of matter, $ \Omega_m $, dominates initially ($ \Omega_m = 1 $) and then starts to decrease as the universe enters into the dark energy-dominated era. Conversely, the density parameter of the quintessence field $ \Omega_{\phi} $ starts to increase. Note that, at the present time ($ z = 0 $), $ \Omega_m \approx 0.3 $ and $ \Omega_{\phi} \approx 0.7 $.

Fig. \ref{eos} shows the evolution of the background equation of state for the quintessence field $ w_{\phi} \equiv \rho_{\phi} / p_{\phi} $. Initially, in the matter-dominated era, the value of $ w_{\phi} $ is locked at $ -1 $. Then, $ w_{\phi} $ starts to depart from $ -1 $ as $ \Omega_{\phi} $ begins to increase. At present time ($ z = 0 $), $ w_{\phi} \approx -0.96 $, which is consistent with the observational data \cite{Ade:2015xua}. Furthermore, the quintessence considered in this work can be classified as a thawing model according to the categorization of the evolution of $ w_{\phi} $ \cite {Caldwell:2005tm, Amendola:2015ksp}.

%%%%%%%%%%%%%%%%%%%%%%%%%%%%%%%%%%%%%%%%%%%%%%%%%%%%%%%%%%%%
\subsection{Evolution of Scalar Perturbations}\label{sec4.2}
%%%%%%%%%%%%%%%%%%%%%%%%%%%%%%%%%%%%%%%%%%%%%%%%%%%%%%%%%%%%

\begin{figure}[t]
 \centering
 \includegraphics[width=\linewidth]{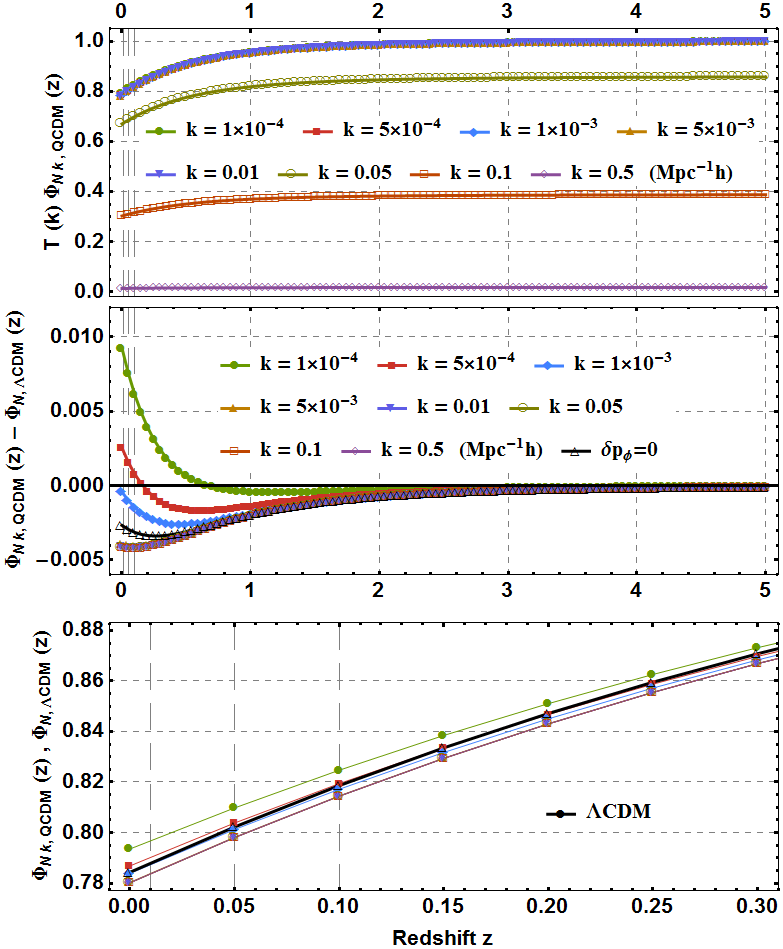}
 \caption{Top: evolution of the weighted gravitational potential $ T(k) \Phi_{Nk}(z) $ at selected scales.  Middle: residue of $ \Phi_{Nk}(z) $ between QCDM and $ \Lambda $CDM, with the additional case $ \delta p_{\phi} = 0 $ in Eq. \eqref{le} (black). Bottom: enlarged plot of $ \Phi_{Nk}(z) $ in QCDM with the same scales as in top panel, compared with $ \Phi_{N}(z) $ in $ \Lambda $CDM (black). The three vertical long-dashed lines correspond to $ z = 0.01 $, $ 0.05 $ and $ 0.1 $, respectively.}\label{Phi}
\end{figure}

Based on current constraints on the amplitude of the curvature power spectrum $ A_{\Phi} $ and on the tensor-to-scalar ratio $ r $ \cite{Ade:2015lrj}, we choose $ 10^9 A_{\Phi} = 2.105 \pm 0.030 $ and $ r = 0.1 $, respectively.
The top panel of Fig. \ref{Phi} shows the evolution of the weighted gravitational potential $ T(k) \Phi_{Nk}(z) $ at selected scales. The gravitational potential at each scale starts to decay when the universe enters into the dark energy-dominant era. Although the curves corresponding to the first five scales, $ k = 10^{-4} $ to $ 0.01 $ $ \textrm{Mpc}^{-1} \textrm{h} $, are almost aligned due to the fact that the transfer function $ T(k) \approx 1 $, they are distinguishable at late times.

The middle panel of Fig. \ref{Phi} shows the residue of $ \Phi_{Nk}(z) $ at each scale between QCDM and $ \Lambda $CDM, with the additional case $ \delta p_{\phi} = 0$ in Eq. \eqref{le}. Note that $ \Phi_N $ has no dependence on $ k $ in this latter case of $ \delta p_{\phi} = 0 $, as well as in $ \Lambda $CDM. The two still deviate from each other, with different rate of decay for $ \Phi_N(z) $ (due to their different background equations of state), leading to a turnover at late times \cite{Jassal:2009gc}.

In the presence of $ \delta p_{\phi} $, a similar trend is observed with a modulation of the QCDM decaying rate depending on scales, as illustrated in the bottom panel of Fig. \ref{Phi}. $ \delta \phi $ is larger at large scales, making the third term in Eq. \eqref{rhop} dominant and bringing a positive deviation to $ \Phi_{Nk}(z) $. The largest deviation is given by the largest scale $ k = 10^{-4} $ $ \textrm{Mpc}^{-1} \textrm{h} $, dominating other modes at all times. On the other hand, the second term in Eq. \eqref{rhop} dominates at small scales, bringing a negative contribution. The curves are convergent at scales smaller than around $ k = 5 \times 10^{-3} $ $ \textrm{Mpc}^{-1} \textrm{h} $, due to the very small variation of this second term with respect to $ k $. In between these two regimes, we can find a scale $ k^\ast \sim 1.6 \times 10^{-3} $ $ \textrm{Mpc}^{-1} \textrm{h} $ for which the variation of $ \Phi_{Nk}(z) $ approaches the case of $ \delta p_{\phi} = 0$ (with still fluctuations around it).
The three vertical long-dashed lines correspond to $ z = 0.01 $, $ 0.05 $ and $ 0.1 $, respectively, at which the luminosity distance power spectrum is calculated.

%%%%%%%%%%%%%%%%%%%%%%%%%%%%%%%%%%%%%%%%%%%%%%%%%%%%%%%%%%%%%
\subsection{Luminosity Distance Power Spectrum}\label{sec4.3}
%%%%%%%%%%%%%%%%%%%%%%%%%%%%%%%%%%%%%%%%%%%%%%%%%%%%%%%%%%%%%

\begin{figure}[t]
 \centering
 \includegraphics[width=\linewidth]{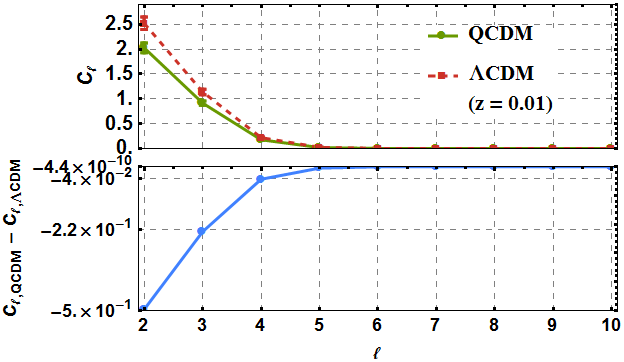}
 \caption{Top: luminosity distance power spectrum $ C_{\ell} $ at $ z= 0.01 $ with integration ranging from $ k = 10^{-4} $ to $ 0.1 $  $ \textrm{Mpc}^{-1} \textrm{h} $ for each $ \ell = 2 $ to $ 10 $. Bottom: residue of the luminosity distance power spectrum between QCDM and $ \Lambda $CDM.}\label{LDPSz0.01}
\end{figure}
\begin{figure}[t]
 \centering
 \includegraphics[width=\linewidth]{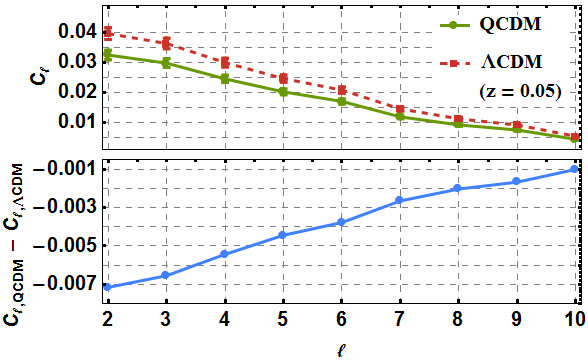}
 \caption{Top: luminosity distance power spectrum $ C_{\ell} $ at $ z= 0.05 $ with integration ranging from $ k = 10^{-4} $ to $ 0.1 $  $ \textrm{Mpc}^{-1} \textrm{h} $ for each $ \ell = 2 $ to $ 10 $. Bottom: residue of the luminosity distance power spectrum between QCDM and $ \Lambda $CDM.}\label{LDPSz0.05}
\end{figure}
\begin{figure}[t]
 \centering
 \includegraphics[width=\linewidth]{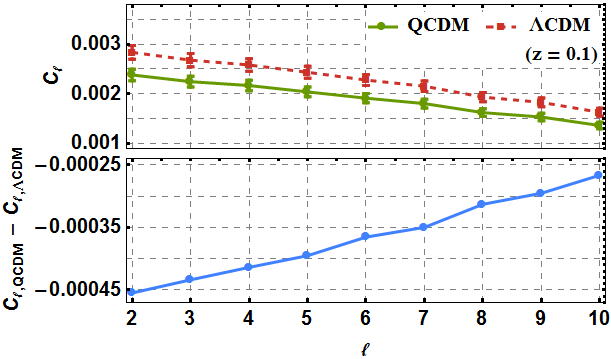}
 \caption{Top: luminosity distance power spectrum $ C_{\ell} $ at $ z= 0.1 $ with integration ranging from $ k = 10^{-4} $ to $ 0.1 $  $ \textrm{Mpc}^{-1} \textrm{h} $ for each $ \ell = 2 $ to $ 10 $. Bottom: residue of the luminosity distance power spectrum between QCDM and $ \Lambda $CDM.}\label{LDPSz0.1}
\end{figure}

In order to calculate the luminosity distance power spectrum efficiently, we construct some artificial fitting functions to mimic $ \Phi_{Nk}(\eta) $ and all integrals over $ \eta $ involved in it, and prepare a look-up table via sampling over the $ ( k , \eta ) $ space. The errors between the mimicked and original functions are less than $ \mathcal{O}(10^{-6}) $.

According to Fig. \ref{Phi}, we set the integration range of $ k $ from $ 10^{-4} $ to $ 0.1 $ $ \textrm{Mpc}^{-1} \textrm{h} $, since the small deviation of $ T(k) \Phi_{Nk}(z) $ for the scales smaller than $ k = 0.1 $ $ \textrm{Mpc}^{-1} \textrm{h} $ will contribute little to the discrimination between QCDM and $ \Lambda $CDM.
The top panels of Fig. \ref{LDPSz0.01}, \ref{LDPSz0.05} and  \ref{LDPSz0.1} show the luminosity distance power spectrum $ C_{\ell} $ as a function of $ \ell $ (from 2 to 10), at $ z = 0.01 $, $ 0.05 $ and $ 0.1 $ respectively, all with integrations ranging from $ k = 10^{-4} $ to $ 0.1 $ $ \textrm{Mpc}^{-1} \textrm{h} $. The error bar comes from the measurement error of $ A_{\Phi} $ \cite{Ade:2015lrj}, which is present in the integration \cite{Bonvin:2005ps}. The bottom panels of Fig. \ref{LDPSz0.01}, \ref{LDPSz0.05} and \ref{LDPSz0.1} show the residue of the luminosity distance power spectrum between QCDM and $ \Lambda $CDM.
This is the main result of this paper. The power in QCDM is smaller than that in $ \Lambda $CDM. The deficit of power is particularly significant for smaller redshift and low multipoles, which is due to the positive deviation of $ \Phi_{Nk}(z) $ for QCDM at small redshit and large scales. The positive deviation at large scales and late times indicates that the $ \Phi_{Nk}(z) $ in QCDM is less decayed compared to that in $ \Lambda $CDM. That is, the photons have to consume more energy to climb out the gravitational potential, leading to the deficit of power.

%%%%%%%%%%%%%%%%%%%%%
\section{Conclusions}
%%%%%%%%%%%%%%%%%%%%%

In this paper, we adopt the quintessence field with the exponential potential as the simplest candidate for the dynamical dark energy model, allowed by the observations, to demonstrate the new approach to distinguish between the dynamical dark energy model and the cosmological constant.  We consider the quantum fluctuations of quintessence field as the physical origin of $ \delta \phi $ itself which is isocurvature perturbation and has fully positive correlation with the primordial curvature perturbation, and its effect on the cosmic expansion as well as on the gravitational potential. We show that the primordial amplitude of $ \delta \phi $ can be related to the tensor-to-scalar ratio $ r $. The residue of gravitational potential between QCDM and $ \Lambda $CDM depends on scales. The largest residue appears at the largest scale of $ k = 10^{-4} $ $ \textrm{Mpc}^{-1} \textrm{h} $ at present time.
The perturbed luminosity distance and its power spectrum were calculated numerically with the evolution of the gravitational potential. 
We find that the smaller the redshift and multipole, the more deficit of power in QCDM compared to that in $ \Lambda $CDM, induced by the effect from quantum fluctuations of quintessence field, which is a foreground effect at large scales and late times. 
Our results are consistent with that in \citep{Moroi:2003pq}, concerning the conclusions of the suppression in low multipoles of CMB temperature anisotropies.

Observationally, this approach can be tested through (future) measurements of the luminosity distance of supernovae \cite{Scolnic:2015eyc, Suzuki:2011hu} (high accuracy but low number) combined with that of galaxies complementary to supernovae \cite{Tully:2016ppz} (large number but low accuracy).
To investigate the nature of dark energy, the luminosity distance power spectrum is indeed an important and useful tool which can be treated as an estimator of how anisotropic the late-time accelerated expansion is induced by the dynamical dark energy field, and as a discriminator between the cosmological constant and a dynamical field that drives the late-time accelerated expansion.
\\

\begin{acknowledgments}
We would like to thank F. Arroja, K. Izumi and An. E. Romano for very useful comments and suggestions, and a special gratitude goes to H. W. Chiang and F. Nugier, who helped to significantly improve the text of this paper.
\end{acknowledgments}

%%%%%%%%%%%%%%%%%%%%%%%%%%%%%%%%%%%%%%%%%%%%%%%%%%%%%%%%%%%%%%%%%%%%%%%%%%%%%%%%%%%%%%%%%
\appendix \section{The Perturbed Luminosity Distance}\label{appa} 
%%%%%%%%%%%%%%%%%%%%%%%%%%%%%%%%%%%%%%%%%%%%%%%%%%%%%%%%%%%%%%%%%%%%%%%%%%%%%%%%%%%%%%%%%

To first order in scalar perturbations, the fully perturbed luminosity distance $ d_L(z_S,\mathbf{n}) $ of a source at redshift $ z_S $ in the radial direction $ \mathbf{n} $ is formulated in \cite{Bonvin:2005ps} as
\begin{widetext}
\begin{align}\label{pld}
  \displaystyle
  d_L(z_S,\mathbf{n}) = & (1 + z_S)(\eta_O - \eta_S)
    \left\{ 1 - \frac{1}{(\eta_O - \eta_S)\mathcal{H}_S} \mathbf{v}_O \cdot \mathbf{n}
            - \left( 1 - \frac{1}{(\eta_O - \eta_S)\mathcal{H}_S} \right) \mathbf{v}_S \cdot \mathbf{n}
    \right.  \nonumber \\
  & - \frac{1}{(\eta_O - \eta_S)\mathcal{H}_S} \Phi_O - \left( 1 - \frac{1}{(\eta_O - \eta_S)\mathcal{H}_S} \right) \Phi_S  \nonumber \\
  & + \frac{2}{\eta_O - \eta_S} \int^{\eta_O}_{\eta_S} d\eta \, \Phi
    + \frac{2}{(\eta_O - \eta_S)\mathcal{H}_S} \int^{\eta_O}_{\eta_S} d\eta \, \Phi'
    - 2 \int^{\eta_O}_{\eta_S} d\eta \, \frac{\eta - \eta_S}{\eta_O - \eta_S} \Phi'
    + \int^{\eta_O}_{\eta_S} d\eta \, \frac{(\eta - \eta_S)(\eta_O - \eta)}{\eta_O - \eta_S} \Phi''  \nonumber \\
  & \left. - \int^{\eta_O}_{\eta_S} d\eta \, \frac{(\eta - \eta_S)(\eta_O - \eta)}{\eta_O - \eta_S} \nabla^2 \Phi
    \right\}.
\end{align}
\end{widetext}
In the above equation, the first term in the brace, the unity, is the background contribution without perturbations. The remaining part of the fist line is attributed to the peculiar motions of the observer and the emitter. The terms in the second line can be regarded as the effects of gravitational redshift. The terms in the third line contain the integrated effects proportional to the line of sight integrals of $ \Phi $ and its time derivative. The last line is the lensing convergence term with $ \nabla^2 \Phi \propto \delta \rho $.

%%%%%%%%%%%%%%%%%%%%%%%%%%%%%%%%%%%%%%%%%%%%%%%%%%%%%%%%%%%%%%%%%%%%%%%%%%%%%%%%%%%%%%%%%
\section{The Components $ C^{(i)}_{\ell} $ of the Luminosity Distance Power Spectrum}\label{appb} 
%%%%%%%%%%%%%%%%%%%%%%%%%%%%%%%%%%%%%%%%%%%%%%%%%%%%%%%%%%%%%%%%%%%%%%%%%%%%%%%%%%%%%%%%%

The $ C^{(i)}_{\ell} $ collects all the contributions to $ C_{\ell} $ which contain integrals of the form $ \int dk k^{i-2} $. For the case when $ z_S = z_{S'} $, the detailed expressions of $ C^{(i)}_{\ell} $ are given by

\begin{widetext}
\begin{align}
  \displaystyle
  C_{\ell}^{(1)} = \frac{2}{\pi} \int \frac{dk}{k} P_{\Phi}(k)
    \left[ \frac{2}{\eta_O - \eta_S} \int^{\eta_O}_{\eta_S} d\eta \, T(k) \Phi_k(\eta) j_{\ell} [k(\eta_O - \eta)]
           - \left( 1 -  \frac{1}{(\eta_O - \eta_S)\mathcal{H}_S} \right) T(k) \Phi_k(\eta_S) j_{\ell}[k(\eta_O - \eta_S)]
    \right]^2,
\end{align}
\begin{align}
  \displaystyle
  C_{\ell}^{(2)} = \frac{-8}{\pi} & \int dk \, P_{\Phi}(k)
    \left[ \frac{1}{4\mathcal{H}_S - \frac{2a''_S}{a_S}}
      \left( 1 - \frac{1}{(\eta_O - \eta_S)\mathcal{H}_S} \right)
      \left( T(k) \Phi_k(\eta_S) + \frac{1}{\mathcal{H}_S} T(k) \Phi'_k(\eta_S) \right) j'_{\ell}[k(\eta_O - \eta_S)]
    \right.  \nonumber \\
  & \displaystyle \left.
    - \frac{1}{(\eta_O - \eta_S)\mathcal{H}_S} \int^{\eta_O}_{\eta_S} d\eta \, T(k) \Phi_k(\eta) j'_{\ell}[k(\eta_O - \eta)]
    + \frac{1}{(\eta_O - \eta_S)} \int^{\eta_O}_{\eta_S} d\eta \int^{\eta}_{\eta_S} d\tilde{\eta} \, 
      T(k) \Phi_k(\tilde{\eta}) j'_{\ell}[k(\eta_O - \tilde{\eta})]
    \right]  \nonumber \\
  & \displaystyle \times
    \left[ \frac{2}{\eta_O - \eta_S} \int^{\eta_O}_{\eta_S} d\eta \, T(k) \Phi_k(\eta) j_{\ell} [k(\eta_O - \eta)]
           - \left( 1 -  \frac{1}{(\eta_O - \eta_S)\mathcal{H}_S} \right) T(k) \Phi_k(\eta_S) j_{\ell}[k(\eta_O - \eta_S)]
    \right],
\end{align}
\begin{align}
  \displaystyle
  C_{\ell}^{(3)} = \frac{8}{\pi} & \int dk \, k P_{\Phi}(k)
    \left[ \frac{1}{4\mathcal{H}_S - \frac{2a''_S}{a_S}}
      \left( 1 - \frac{1}{(\eta_O - \eta_S)\mathcal{H}_S} \right)
      \left( T(k) \Phi_k(\eta_S) + \frac{1}{\mathcal{H}_S} T(k) \Phi'_k(\eta_S) \right) j'_{\ell}[k(\eta_O - \eta_S)]
    \right.  \nonumber \\
  & \displaystyle \left.
    - \frac{1}{(\eta_O - \eta_S)\mathcal{H}_S} \int^{\eta_O}_{\eta_S} d\eta \, T(k) \Phi_k(\eta) j'_{\ell}[k(\eta_O - \eta)]
    + \frac{1}{(\eta_O - \eta_S)} \int^{\eta_O}_{\eta_S} d\eta \int^{\eta}_{\eta_S} d\tilde{\eta} \, 
      T(k) \Phi_k(\tilde{\eta}) j'_{\ell}[k(\eta_o - \tilde{\eta})]
    \right]^2  \nonumber \\
  & \displaystyle + \frac{4}{\pi} \int dk \, k P_{\Phi}(k)
    \left[ \frac{1}{(\eta_O - \eta_S)} \int^{\eta_O}_{\eta_S} d\eta \int^{\eta}_{\eta_S} d\tilde{\eta} \, 
      (\tilde{\eta} - \eta_s) T(k) \Phi_k(\tilde{\eta})
      \left( j_{\ell}[k(\eta_O - \tilde{\eta})] + j''_{\ell}[k(\eta_O - \tilde{\eta})] \right)
    \right]  \nonumber \\
  & \displaystyle \times
    \left[ \frac{2}{\eta_O - \eta_S} \int^{\eta_O}_{\eta_S} d\eta \, T(k) \Phi_k(\eta) j_{\ell} [k(\eta_O - \eta)]
          - \left( 1 -  \frac{1}{(\eta_O - \eta_S)\mathcal{H}_S} \right) T(k) \Phi_k(\eta_S) j_{\ell}[k(\eta_O - \eta_S)]
    \right],
\end{align}
\begin{align}
  \displaystyle
  C_{\ell}^{(4)} = \frac{-8}{\pi} & \int dk \, k^2 P_{\Phi}(k)
      \left[ \frac{1}{4\mathcal{H}_S - \frac{2a''_S}{a_S}}
      \left( 1 - \frac{1}{(\eta_O - \eta_S)\mathcal{H}_S} \right)
      \left( T(k) \Phi_k(\eta_S) + \frac{1}{\mathcal{H}_S} T(k) \Phi'_k(\eta_S) \right) j'_{\ell}[k(\eta_O - \eta_S)]
    \right.  \nonumber \\
  & \displaystyle \left.
    - \frac{1}{(\eta_O - \eta_S)\mathcal{H}_S} \int^{\eta_O}_{\eta_S} d\eta \, T(k) \Phi_k(\eta) j'_{\ell}[k(\eta_O - \eta)]
    + \frac{1}{(\eta_O - \eta_S)} \int^{\eta_O}_{\eta_S} d\eta \int^{\eta}_{\eta_S} d\tilde{\eta} \, 
      T(k) \Phi_k(\tilde{\eta}) j'_{\ell}[k(\eta_O - \tilde{\eta})]
    \right]  \nonumber \\
  & \displaystyle \times
    \left[ \frac{1}{(\eta_O - \eta_S)} \int^{\eta_O}_{\eta_S} d\eta \int^{\eta}_{\eta_S} d\tilde{\eta} \, 
      (\tilde{\eta} - \eta_S) T(k) \Phi_k(\tilde{\eta})
      \left( j_{\ell}[k(\eta_O - \tilde{\eta})] + j''_{\ell}[k(\eta_O - \tilde{\eta})] \right)
    \right],
\end{align}
\begin{align}
  \displaystyle
  C_{\ell}^{(5)} = \frac{2}{\pi} & \int dk \, k^3 P_{\Phi}(k)
    \left[ \frac{1}{(\eta_O - \eta_S)} \int^{\eta_O}_{\eta_S} d\eta \int^{\eta}_{\eta_S} d\tilde{\eta} \, 
      (\tilde{\eta} - \eta_S) T(k) \Phi_k(\tilde{\eta})
      \left( j_{\ell}[k(\eta_O - \tilde{\eta})] + j''_{\ell}[k(\eta_O - \tilde{\eta})] \right)
    \right]^2,
\end{align}
\end{widetext}
where, for simplicity, we assume that the primordial curvature power spectrum $ P_{\Phi}(k) $ is scale invariant, $ P_{\Phi}(k) \simeq A_{\Phi} (k \eta_0)^{n-1} $ with $ n \simeq 1 $.

\bibliographystyle{unsrt}
\bibliography{reference}

\end{document}